\documentclass[aps,amsfonts,pre,preprint,showpacs,nobibnotes,nofootinbib]{revtex4}
\usepackage{graphicx}
\begin{document}
\newcommand{\beq}{\begin{equation}}
\newcommand{\eeq}{\end{equation}}
\newcommand{\barr}{\begin{eqnarray}}
\newcommand{\earr}{\end{eqnarray}}
\newcommand{\avg}[1]{\left< #1 \right>}
\def\figwidth{7.5cm}
\newcommand{\Mfunction}[1]{#1}
\def\cH{{\cal H}}
\def\cT{{\cal T}}
\def\cP{{\cal P}}
\def\cD{{\cal D}}
\def\cG{{\cal G}}
\def\cV{{\cal V}}
\def\cP{{\cal P}}
\def\cF{{\cal F}}
\def\cL{{\cal L}}
\def\cR{{\cal R}}
\def\cU{{\cal U}}
\def\cS{{\cal S}}
\def\cO{{\cal O}}
\def\cE{{\cal E}}
\def\bfA{{\bf A}}
\def\bfG{{\bf G}}
\def\bfn{{\bf n}}
\def\bfr{{\bf r}}
\def\bfV{{\bf V}}
\def\bft{{\bf t}}
\def\bfM{{\bf M}}
\def\bfS{{\bf S}}
\def\bfP{{\bf P}}
\def\vecj{\vec{j}}
\def\vnabla{\vec{\nabla}}
\def\vS{\vec{S}}
\def\vn{\vec{n}}
\def\vk{\vec{k}}
\def\vx{\vec{x}}
\def\svn{{\vec{n}}}
\def\bra#1{\langle #1 |}
\def\ket#1{| #1 \rangle}
\newcommand{\p}{\partial}
\def\coltwovector#1#2{\left({#1\atop#2}\right)}
\def\upp{\coltwovector10}
\def\downn{\coltwovector01}
\def\Ord#1{{\cal O}\left( #1\right)}
\def\bmp{\mbox{\boldmath $p$}}
\def\rhobar{\bar{\rho}}
\renewcommand{\Re}{{\rm Re}}
\renewcommand{\Im}{{\rm Im}}
\title{On the connection between Random Waves and Quantum Fields.
\\
Duality between nodal lines statistic and the Casimir energy
}\author{A.\ Scardicchio}\email{scardicc@mit.edu}
\affiliation{Center for Theoretical Physics, \\ Laboratory for Nuclear Sciences and Physics Department\\
Massachusetts Institute
  of Technology \\ Cambridge, MA 02139, USA}
\begin{abstract}
    \noindent Using the statistical description common to random waves and quantum fields
    we show how the probability of
    having a nodal line close to a (translationally symmetric) reference curve $\gamma$ is
    related to the Casimir energy of an appropriate configuration
    of conductors.
\end{abstract}
 \vspace*{-\bigskipamount}
\preprint{MIT-CTP-3626} \maketitle \setcounter{equation}{0}

\section{Introduction}

Random waves (RW) have been an object of interest for their
statistical properties both in wave mechanics and statistical
mechanics. In wave mechanics they turned out to be an incredibly
interesting and rich scenario for studying the statistics of
topological properties, like phase singularities \cite{Berry1}. In
optics they made a good statistical model for speckle patterns
\cite{Halperin1} in laser beams. In quantum mechanics they have
been studied \cite{Berry0,Smilansky} in connection with
semiclassical wave functions in chaotic billiards. In statistical
mechanics their properties have been put in connection with the
statistic of defects and vortices \cite{Halperin2} and an
interesting duality with a percolation problem has been put
forward recently \cite{Bogomolny}.

The purpose of this paper is to exploit in a new direction the
description of RW in terms of quantum field theory
(QFT).\footnote{The statistical mechanics description of RW is
related to this via the usual QFT-statistical mechanics duality
(Wick rotation). In this sense this description is already
contained in the work of B.~Halperin in \cite{Halperin1} who
employed it to study the statistic of vortices and defects.} I
will make a quantitative connection between the ground state
energy (or Casimir energy) of a scalar field in a given
configuration of semi-penetrable conductors in $d-1$ space
dimensions and the probability of having a certain configuration
of nodal lines in $d$ dimensions.

This paper is far from being exhaustive or self-contained. I will
briefly introduce the concept of random wave referring the reader
to the existing literature for their statistical properties, in
particular the properties of their nodal lines. I will then
rephrase the concept of Casimir energy for a scalar field in a
static background (a toy model for QED where the static background
models the conductors) in a language closer to that of random
waves statistics. I will then point out the connection between
Casimir energy in this background and probability of having a
certain nodal line configuration. Finally I will draw some
consequences on the nodal line probability and comment on the
possible extensions on which further work is needed.

\section{Random Waves}

An isotropic random wave (RW) in $d$ dimensions is the random
function defined on a subset of $\mathbb{R}^d$ (we will not use
any particular notation for vectors but there is little room for
confusion) as
\beq
\label{eq:rw1}
\phi(x)=\sum_{j=1}^J\sqrt{\frac{2}{J}}\epsilon(k_j)\cos(k_j
x+\delta_j)
\eeq
where the phases $\delta_j$ are uniformly distributed in
$[0,2\pi)$ and the vectors $k_j$ are random variables as well. We
will assume isotropy of $\epsilon$, \emph{i.e.} $\epsilon(k)$ is
an even, analytic function of the length of the vector $k$.

For any finite $J$ the moments $\avg{\phi(x_1)...\phi(x_n)}$ are
not factorizable, but in the limit $J\to\infty$ Wick theorem holds
\cite{Berry1} (among other things one requires the existence and
finiteness of at least the second moment, \emph{i.e.}
$\avg{\phi^2(x)}<\infty $):
\beq
\avg{\phi(x_1)...\phi(x_{2n})}=\sum_{{\rm
Contractions}}\avg{\phi(x_i)\phi(x_j)}...\avg{\phi(x_k)\phi(x_l)}.
\eeq
In the following we will hence always assume the limit
$J\to\infty$ is taken.

Wick's theorem is equivalent to saying that the statistical
properties of RW can be described by a Gaussian probability
functional
\beq
\label{eq:Ph}
P[\phi]=\frac{1}{Z}\exp\left(-\frac{1}{2}\int
d^dxd^dx'\phi(x')h(x',x)\phi(x)\right),
\eeq
where $Z$ is a normalization constant and $h(x',x)=h(|x'-x|)$ for
isotropic RW. From this probability functional the reader could
already recognize the usual set-up of the statistical mechanics of
a non-interacting real field $\phi$. The function $h$ is
determined by the spectrum $\epsilon(k)$ (and vice versa). We will
now determine their connection.

To this purpose is convenient to pass to the Fourier components of
the field $\phi_k=\int d^dxe^{ikx}\phi(x)$ and define $h(k)$
through
\beq
\int d^dxd^dx'e^{ikx-ik'x'}h(x',x)=(2\pi)^d\delta^{(d)}(k'-k)h(k).
\eeq
In terms of $\phi_k$ the probability functional is
\beq
\label{eq:Phk}
P[\phi]=\frac{1}{Z}\exp\left(-\frac{1}{2}\int\frac{d^dk}{(2\pi)^d}h(k)|\phi_k|^2\right).
\eeq
The limiting Gaussian probability functional (\ref{eq:Ph}) or
(\ref{eq:Phk}) can describes the statistical properties of
(\ref{eq:rw1}) if we choose the spectrum $\epsilon(k)$ as
\beq
\label{eq:spectrdual}
\lim_{J\to\infty}\frac{1}{J}\epsilon^2(k)=\frac{1}{h(k)}\frac{d^dk}{(2\pi)^d}
\eeq
which means that in the limit $J\to\infty$ the sum over $k_j$ must
be substituted by the integral in $d^dk$ whose measure is given by
right-hand side of (\ref{eq:spectrdual}). This is the promised
connection between $h(x)$ and $\epsilon(k)$.
In this way when $J\to\infty$ the propagator $G$ tends to
\beq
G(x,0)\equiv\avg{\phi(x)\phi(0)}=\lim_{J\to\infty}\sum_{j=1}^J\frac{1}{J}\epsilon^2(k_j)\cos(k_jx)
=\int\frac{d^dk}{(2\pi)^d}\frac{1}{h(k)}e^{ikx},
\eeq
where we used the fact that $\epsilon$ is even in $k$ to
substitute $e^{ikx}$ for $\cos(kx)$.

There are at least two `natural' choices for the spectrum $h(k)$:
\begin{itemize}

\item The \emph{scalar field} spectrum
$1/h(k)=\theta(\Lambda-|k|)/(k^2+m^2)$ where one has to introduce
the cutoff $\Lambda$ to ensure the finiteness of
$G(x,x)=\avg{\phi^2(x)}$.

\item The very singular \emph{monochromatic} spectrum $h(k)$, such
that $1/(2\pi h(k))=\delta(|k|-K)$. This last choice gives
$G(x,0)=J_0(Kx)$ which is a statistical model for the solutions of
the Schr\"odinger equation $-\Delta \psi=K^2\psi$ in chaotic
billiards.

\end{itemize}

\section{Casimir Energy and Nodal Lines}

We now turn to the main point of this paper: the connection
between nodal lines properties and Casimir energy. For simplicity
at the moment we assume $d=2$, the generalization to other $d$
will be straightforward.

Following \cite{Smilansky} we introduce the functional
\beq
X_\gamma[\phi]=\frac{1}{2}\int_\gamma ds \phi^2(x(s)).
\eeq
where the integral is defined over the reference line
$\gamma=\{x(s)|s\ \epsilon[0,\ell] \}$ and parameterized with the
length of the line itself, $s$. For any given reference curve
$\gamma$, $X_\gamma[\phi]$ is a random variable whose generating
function $S_\gamma(\lambda)$ is defined as
\beq
\label{def:S}
S_\gamma(\lambda)\equiv\avg{e^{-\lambda X_\gamma[\phi]}}=\int\cD
\phi\ P[\phi]\ e^{-\frac{1}{2}\lambda\int_\gamma ds\
\phi^2(x(s))}.
\eeq
It has been shown in \cite{Smilansky} that $S_\gamma(\lambda)$ can
be interpreted approximately as the probability of having a nodal
line in the tube of radius
$r=(\avg{(\nabla\phi)^2}\lambda)^{-1/3}$ built around the
reference curve $\gamma$ (in $d$ dimensions $1/3$ gets substitutes
by $1/(d+1)$). Notice that the radius $r$ of the tube goes to zero
when $\lambda\to\infty$. The approximation allowing us to
interpret $S_\gamma$ as the probability of having a nodal line
relies mainly on a mean-field approximation where $\phi^2/(\nabla
\phi^2)\to \phi^2/\avg{(\nabla \phi)^2}$ as discussed in
\cite{Smilansky}. It is not easy to estabilish the limits of this
approximation so we will adopt it as a working hypothesis and we
will see later a situation in which it possibly fails. From now on
we will simply say that $S_\gamma$ is `the probability to have a
nodal curve $\gamma$' without referring to the tube radius $r$ or
the approximation within which this interpretation has been
derived.

Let us write in (\ref{def:S}) $P[\phi]$ explicitly inside the
probability functional
\beq
S_\gamma(\lambda)=\int
\cD\phi\frac{1}{Z}\exp\left(-\frac{1}{2}\int d^2xd^2x'\
\phi(x')(h(x',x)+\delta^{(2)}(x'-x)V(x))\phi(x)\right),
\eeq
where we have defined
\beq
V(x)=\lambda\int_\gamma ds\ \delta^{(2)}(x-x(s)).
\eeq

Let us now specialize the problem in two ways:
\begin{itemize}
\item Choose $h(x',x)$ to mimic a scalar field, with a cutoff
$\Lambda$ intended in all the momentum integrals
\beq
h(x',x)=\delta^{(2)}(x'-x)(-\Delta+m^2).
\eeq
\item Consider a random wave in the strip $[0,T]\times\mathbb{R}$.
Denote the two cartesian coordinates in the plane as $x_0,x_1$ so
$0\leq x_0\leq T$ and $x_1\in \mathbb{R}$. Choose the reference
line $\gamma$ as made of $n\geq 1$ disconnected lines parallel to
the $x_0$ axis and intersecting the $x_1$ axis at the points
$\{a_1,...,a_n\}$
\beq
\gamma=\{a_1,...,a_n\}\times[0,T].
\eeq
\end{itemize}

With this assumptions the final expression for the generating
function $S_\gamma(\lambda)$ is then
\beq
\label{eq:Slambda}
S_\gamma(\lambda)=\frac{1}{Z}\int
\cD\phi\exp\left(-\frac{1}{2}\int_{[0,T]\times \mathbb{R}} d^2x\
\phi(x)(-\Delta+m^2+V(x))\phi(x)\right)
\eeq
where $\Delta=\partial^2/\partial x_0^2+\partial^2/\partial
x_1^2$. This expression itself is reminiscent of two intertwined
concepts in QFT and statistical field theory: the Casimir energy
$\cE$ in the first and the free energy $F$ in the second. The
connection with the latter is evident, without any need for formal
manipulations, $F=-\log(S_\gamma(\lambda))$. The connection with
the Casimir energy becomes evident as well if we perform a
clockwise (inverse) Wick rotation in the $x_0$ coordinate, $x_0\to
it$. Then (\ref{eq:Slambda}) becomes
\beq
\frac{1}{Z}\int \cD\phi\exp\left(i\frac{1}{2}\int_{[0,T]\times
\mathbb{R}} dtdx_1\
\phi(t,x_1)(-\partial^2-m^2-V(x_1))\phi(t,x_1)\right)=e^{-i
\cE_\gamma T}.
\eeq
Here $\partial^2=\partial^2/\partial t^2-\partial^2/\partial
x_1^2$ and $\cE_\gamma$ the Casimir energy in the background $V$
and
\beq
V(y)=\sum_{i=1}^n\delta(a_i-y).
\eeq

We can now establish the promised connection between the
generating functional $S_\gamma(\lambda)$ and the Casimir energy
$\cE_\gamma$ of the corresponding background as
\beq
S_\gamma(\lambda)=e^{-T\cE_\gamma}.
\eeq

In words: \emph{The probability of having a (translationally
symmetric) nodal line $\gamma$ in a random wave ensemble is
related to the Casimir energy of a configuration of conductors
given by a constant-time section of $\gamma$.}

The generalization to $d$ dimension is easily obtained (and is
already understood in the previous paragraph). Since a nodal
hypersurface has codimension 1 so does its constant-$x_0$ section.
Hence the problem maps to the Casimir energy of codimension 1
surfaces. The dual Casimir problem then is the usual problem of
penetrable, codimension 1 surfaces (see \cite{cosmic} for the case
with arbitrary codimension). For example, the $d=4$ case maps into
the $\mathbb{R}^3$ Casimir problem with penetrable 2-dimensional
surfaces. The limit $\lambda\to\infty$ is the limit of perfect
conductors (or Dirichlet limit, because $\phi=0$ on the conducting
surfaces).

In the rest of the paper we will use this duality to make
statements on the nodal line statistic from the knowledge of the
properties of Casimir energy.

\section{Applications}

Let us start with a well known problem of a Casimir energy
calculation: the presence of various divergencies, when taking
$\Lambda,\ \lambda\to\infty$. We will now discuss the
interpretation of these divergencies for the nodal lines
probability.

A volume divergence $\propto V\Lambda^{d+1}$ (divergent when
$\Lambda\to\infty$) is removed by the factor of $1/Z$ in our
definition of $P[\phi]$. This term is however independent of the
presence and shape of $\gamma$. In QFT it would represent a
cosmological constant term. For what concerns the divergencies
arising when $\lambda\to\infty$ (the so-called Dirichlet limit)
let us recall that the tube radius $r$ built around the reference
line $\gamma$, goes to $0$ when $\lambda\to\infty$. It is then
natural that the probability $S_\gamma(\lambda)$ of having a nodal
line within a distance $r$ from $\gamma$ goes to zero when
$\lambda\to\infty$. This reflects in the fact that $\cE\to+\infty$
when $\lambda\to\infty$.

We also know that the interaction energy (the one that depends on
the distance between the bodies) remains finite when
$\Lambda,\lambda\to\infty$. This means that there are some
properties of the nodal lines, connected with the interaction part
of $\cE$, which are well-defined also when the tube radius $r\to
0$ and the cutoff goes to infinity. In order to identify them we
must define a quantity which stays finite in this limit. Led by
the intuition about the Casimir energy of rigid conductors we
recognize that this problem is related to the removal of the
self-energy for rigid, disconnected bodies. Suppose then our curve
$\gamma$ is composed of two disconnected pieces
$\gamma=\gamma_1\cup\gamma_2$ (we always require them to be
straight and both parallel in the $x_0$ direction). Their Casimir
energy can be written as
\beq
\cE_\gamma=\cE_{\gamma_1}+\cE_{\gamma_2}+\cE_{\rm int}
\eeq
where the first two terms are independent on the distance between
the nodal lines and the last term goes to zero when the distance
between the two curves goes to infinity (this can be taken as a
definition of $\cE_{\rm int}$). The terms $\cE_{\gamma_{1,2}}$ are
the energies of isolated plates.

Let us define the quantity $\cP$ as the ratio of the probability
of having $\gamma_1\cup\gamma_2$ and the probability of having
both $\gamma_1$ \emph{and} $\gamma_2$ independently of each other:
\beq
\cP=\frac{S_{\gamma_1+\gamma_2}(\lambda)}{S_{\gamma_1}(\lambda)S_{\gamma_2}(\lambda)}=e^{-T\cE_{\rm
int}}.
\eeq

The interpretation of $\cP$ is the following: if $\cP>1$ ($\cP<1$)
then it is easier (more difficult) to find a nodal line $\gamma_2$
if another line $\gamma_1$ is present.

The interaction energy $\cE_{\rm int}$ is always finite (even when
$m\to 0$ and/or $\Lambda,\lambda\to \infty$) and hence so is
$\cP$. Moreover we know from quantum field theory that $\cE_{\rm
int}<0$ and that it increases when the nodal lines are pulled
apart. Hence we can say that the presence of a nodal line
$\gamma_1$ makes it easier for another nodal line to be born.
Hence in this case nodal lines \emph{induce} other nodal lines in
their vicinity.

The choice of the scalar field spectrum allows us to use all the
machinery of QFT (including the Hamiltonian formulation) to
calculate the Casimir energy $\cE_{\rm int}$. Depending on the
value of $\lambda,\ \Lambda,\ m$ and the distance between the
nodal lines $a$, we can use a weak or a strong coupling
approximation for $\cE_{\rm int}$. Since $\lambda$ has dimension
$\ell^{d-1}$ ($\ell$ is a length scale) the relevant dimensionless
parameter is $\epsilon=a\lambda^{1/(d-1)}$. Moreover assuming
$\Lambda\gg m$, $\avg{(\nabla\phi)^2}\sim \Lambda^2$ we have
$\lambda\sim 1/r^{d+1}\Lambda^2$. By choosing $a,r,\Lambda$ we
have $\epsilon\ll 1$ for weak coupling and $\epsilon\gg 1$ for the
strong coupling regime.

We will now make some explicit sample calculation in these two
regimes.

In the \emph{weak coupling} regime can use a Feynman diagram
expansion \cite{Jaffe} (one must use the Euclidean cutoff
$\Lambda$ on the $k$ integrals) for the Casimir energy $\cE$
\beq
\label{eq:CEexp}
\cE=\lambda \cE_1+\lambda^2 \cE_2+...
\eeq
where $\cE_1$ is given by the tadpole diagram, $\cE_2$ is given by
the 2 legs diagram and so on. We will now calculate the first two
terms of the series (\ref{eq:CEexp}) showing that $\cE_1$ drops
between numerator and denominator in $\cP$ and then calculating
the first correction to $\cP$, i.e.\ $\cE_2$. We will also show
that $\cE_2<0$, which implies $\cP>1$, in a region of order $1/m$
around any nodal line.

For the tadpole diagram we have
\beq
\cE_1=\int d^{d-1}x V(x)\avg{\phi^2(x)}=\avg{\phi^2(0)}\int
d^{d-1}x V(x)
\eeq
or in Fourier space
\beq
\cE_1=\int\frac{d^{d-1}k}{(2\pi)^{d-1}}V(k)\int\frac{d^{d}q}{(2\pi)^{d}}\frac{1}{q^2+m^2},
\eeq
where $V(k)$ is the Fourier transform with respect to the $d-1$
spatial dimensions
\beq
V(k)\equiv\int d^{d-1}xV(x)e^{ikx}
\eeq
so that in $d=2$ and with $V(x)=\delta(x)+\delta(x-a)$ we have
\beq
V(k)=1+e^{ika}.
\eeq
Since $\int dx V(x)$ does not depend on $a$, the tadpole diagram
does not contribute to the interaction energy $\cE_{\rm int}$ and
hence does not contribute to $\cP$. We must then go to the next
diagram, the one with two legs to find the first non zero
correction to $\cE_{\rm int}$. The two-legs diagram contribution
can be written as
\beq
\cE_2=-\int
\frac{d^{d-1}k}{(2\pi)^{d-1}}V(k)V(-k)\int\frac{d^dq}{(2\pi)^d}\frac{1}{(q+k)^2+m^2}\frac{1}{q^2+m^2}.
\eeq
It contains an $a$-dependent interaction term. To calculate this
$a$-dependent term we can send $\Lambda\to\infty$ (for $d=2$ we
can take this limit safely) and by means of the usual technology
for handling Feynman diagrams we find
\beq
\cE_2=-\frac{1}{2\pi}\int_0^1 dx\int_{-\infty}^\infty
\frac{dk}{2\pi}e^{ika}\frac{1}{m^2+x(1-x)k^2}.
\eeq
Performing the integrals gives
\beq
\cE_2=-\frac{1}{2m}\left(1-\Phi(2\sqrt{ma})\right),
\eeq
where $\Phi$ is the error function. Then $\cP$ can be written, to
this order in $\lambda$, as
\beq
\cP=e^{T\frac{\lambda^2}{2m}(1-\Phi(2\sqrt{ma}))}.
\eeq
As we said before $\cP$ decreases when $a$ increases. Moreover for
$a\gg 1/m $ we can do an asymptotic expansion for $\Phi$ finding
\beq
\cP\simeq \exp\left(T\frac{\lambda^2}{4m\sqrt{am}}e^{-4am}\right),
\eeq
so $\cP\simeq 1$ effectively for $a\gg 1/m$.

The strong coupling limit has to be tackled with different,
non-perturbative techniques.

The $d=2$ case can also be solved exactly for any number $n$ of
parallel nodal lines by using the techniques in \cite{cosmic}. The
resulting exact expression for $n\geq 3$ is too cumbersome to be
presented here and we refer the reader to \cite{cosmic} for
details.

Two parallel nodal lines separated by a distance $a$ in the limit
$\Lambda\to\infty$ are dual to the problem of two points in 1
space dimension at a distance $a$. The Casimir energy for this
configuration is:
\beq
\label{eq:EN1delta1d}
\cE_{\rm int}=\frac{1}{4\pi}\int_0^\infty
\frac{dE}{\sqrt{E}}\ln\left(1-\frac{e^{-2a\sqrt{E+m^2}}}{(1+\frac{2}{\lambda}\sqrt{E+m^2})^2}\right).
\eeq

We can use this formula to make some predictions about $\cP$. To
begin we know that $\cE_{\rm int}<0$ and that it has a minimum at
$a=0$ as $\cE_{\rm int}(a=0,\Lambda\gg \lambda)\simeq-\lambda
\log(2)/2\pi$ (here we put for simplicity $m=0$). It can be proved
that this is also equal to $\cE_{\rm single}(2\lambda)-2\cE_{\rm
single}(\lambda)$ (where $\cE_{\rm single}$ is the energy of a
single delta function when $\Lambda\to\infty$), which appeals to
intuition since at $a=0$ we are just superposing two delta
functions to create a delta function with double strength. If
$m>0$ it can be proved that $\cE\propto \exp(-2ma)$ and hence
again $\cP\simeq 1$ when $a\gg 1/m$. In any case we can say that
$\cP$ decreases when $a$ increases.

A difficulty must be noticed here, concerning how far one can push
the interpretation of $S_\gamma$ as the probability of having a
nodal line $\gamma$. Reasoning like in \cite{Berry4}, assuming
Dirichlet boundary conditions on a line $\gamma$ intersecting the
$x_1$ axis at say $x_1=0$ we have then to expand our RW in series
of $\sin(k_j x_1)$ (with random coefficients). Reasonably the
subset of RW that has a nodal line on $\gamma$ should be
expandible in this basis as well. If moreover our spectrum is cut
off at $\Lambda$ then one expects that for $a\ll \pi/\Lambda$ one
should find much fewer nodal lines (the first zero of $\sin z$ is
at $z=\pi$). In fact a similar phenomenon is found in
\cite{Berry4} for the monochromatic spectrum. The nodal line
length density normalized to its asymptotic value goes to $\sim
0.5$ for $x_1=0$. However increasing $x_1$ the nodal line length
density suddenly increases to a value higher than the asymptotic
value and then relaxes, oscillating, to 1. In analogy our quantity
$\cP$ should then start from a value $<1$ at $a=0$, increase in a
region $1/\Lambda$ to a value $\cP>1$ and then relax to $\cP=1$.
Evidently the first, $\Ord{1/\Lambda}$ region is not captured by
our analysis, while the second one is. This, as we said in the
discussion after Eq.\ (\ref{def:S}), can possibly be traced back
to the failure of the `mean field' approximation that was used to
link $S_\gamma$ with the true probability of finding a nodal line
\cite{Smilansky}. We have hence learned that we must assume $a\gg
1/\Lambda$ for our results to hold. Equation (\ref{eq:EN1delta1d})
for $a\gg 1/\Lambda, 1/\lambda,$ and $m=0$ gives
\beq
\cE_{\rm int}=-\frac{\pi}{24a},
\eeq
yielding
\beq
\cP=e^{\pi T/24 a}.
\eeq

The higher dimension ($d>2$) case cannot be solved in general, due
to its strong geometry dependence. The constant time section of
$\gamma$ can be any hypersurface representing disconnected
conductors in space. The Casimir problem is the most generic one
and we do not posses an efficient way of solving it. We know
however how to solve the case of parallel, large (actually,
infinite) $d-2$ hyper-planes (lines in $d=3$, planes in $d=4$
etc.). The result for $m=0,\ \lambda\to\infty$ is
\beq
\cE_{\rm int}\propto -\frac{S}{a^{d-1}}
\eeq
where $S$ is the $d-2$ dimensional area of the hyper-planes, $a$
their separation and the proportionality constant depends on $d$.

One of the main problems of Casimir physics is to find effective
(analytical or numerical) ways of calculating the Casimir energy
for arbitrary configurations of perfect conductors. Despite recent
developments \cite{Optical, Gies} this problem escapes analytical
solution for all but the simple parallel plates case. We then
expect to gain some insights from the other side of the duality,
namely the nodal lines distributions.

\section{Extensions and further developments}

\emph{Extension to different spectra}. It would be interesting to
know how much of what we said in this paper, based on the scalar
field spectrum, is valid for other spectra (like the monochromatic
spectrum). The monochromatic, as well as other kinds of isotropic
spectra cannot be modelled by an Hamiltonian field theory, even
thought their probability functional is gaussian. The high degree
of non-locality of these spectra implies that the free energy $F$
is not extensive. Hence we could not even define a $T$-independent
quantity like the Casimir energy $\cE$. It is hence of great
interest for a field theorist to grasp some of the properties of
these generalized free QFTs in terms of some, more intuitive
perhaps, statistical properties of random waves.

\emph{Extension to codimension $>1$}. Generically nodal lines of
real fields have codimension 1 (lines in the plane, etc.) because
they are defined by a single condition, namely $\phi(x)=0$.
Codimension 2 or higher nodal lines are non-generic and have
extremely low probability of occurring. For example the
probability of having $\phi(x)=0$ at an isolated point requires
both $\phi(x)=0$ and $|\nabla\phi(x)|=0$ at the same point. This
is \emph{extremely} unlikely in the sense that it has measure 0,
and would never show up in a Montecarlo simulation. We know in
fact from \cite{cosmic} that conductors of codimension 2 and
higher cannot be defined with $\lambda>0$. They \emph{must} be
defined as a limit $\lambda\to 0^-$. However the generating
functional $S_\gamma(\lambda)$ is not well-defined for
$\lambda<0$. It diverges badly. Actually, since
$S_\gamma(\lambda)=\avg{e^{-\lambda X_\gamma}}$, for $\lambda<0$
it is finite and only if the probability distribution of
$X_\gamma$ decays at infinity faster than $e^{\lambda X_\gamma}$.
It turns out that one can take the limits ($\lambda\to 0^-$ and
shrinking $\gamma$ to codimension $>1$) in such a way that this
divergence and the infinitesimal probability of a codimension $>1$
nodal line occurring compensate, giving a finite value for
$S_\gamma$.

\emph{Extension to complex fields and phase singularities.} A
nodal line of a complex field is a more interesting object than
that of a real field \cite{Berry1}. Complex field nodal lines are
phase singularities whose strength can be interpreted as a
topological charge \cite{Halperin2}. Various correlation functions
of this charges have been calculated by means of the Gaussian
field technology. It would be interesting to see what the Casimir
energy analogy has to say on these objects.

\emph{Numerics}. One of the main reasons this duality is
interesting is that it could lead to a more efficient numerical
algorithm for computing Casimir energies of conductors of
arbitrary shape. However this issue is beyond the scope of this
paper and we leave them for future work.

\section{Conclusions}

We have shown that there is a dual description of random waves in
terms of quantum field theory. In particular we put forward and
started the exploration of the duality between the probability of
having a nodal line close to a given disconnected reference curve
and the Casimir energy of a configuration of conductors.

We used this duality to infer some properties of the distribution
of nodal lines and we proved that, for the scalar field spectrum,
nodal lines induce other nodal lines in their proximity. This last
statement just follows from the attractive nature of Casimir
interactions.

This duality can be used in the other direction to gain
information on the Casimir energy of an arbitrary configuration of
conductors from the statistical properties of the nodal line.

\section{Acknowledgments}

I would like to thank B.~Halperin and R.~Jaffe for discussions.
This work has been supported in part by the U.S.~Department of
Energy (D.O.E.) under cooperative research
agreement~\#DE-FC02-94ER40818.

\end{document}